\documentclass[11pt]{article}
\pdfoutput=1
  \usepackage{amssymb}
    \usepackage{graphicx}
\newcommand{\sect}[1]{\setcounter{equation}{0}\section{#1}}
\renewcommand{\theequation}{\arabic{section}.\arabic{equation}}

\def\am{angular momentum~}

\def\av{angular velocity~}

\def\as{asymptotically~}

\def\ba{\begin{eqnarray}}

\def\be{\begin{equation}}
\def\bi{\bibitem}
\def\bm{\boldmath}

\def\Bar {\overline}

\def\Bg{\Bar g}

\def\cd{\cdot}

\def\co{coordinate~}
\def\coo{coordinates~}
\def\cy{cylindrical~}

\def\CC{{\cal C}}

\def\dga{\dot\ga}
\def\di{\partial}

\def\dom{\dot\om}

\def\dW{\dot W}
 
\def\ea{\end{eqnarray}} 
\def\ee{\end{equation}}
\def\eee{equation~}
\def\eeee{equations~}
\def\em{energy-momentum~}

\def\ep{\epsilon} 
\def\eq{\equiv~}

\def\ER{Einstein-Rosen~}
\def\fr{\frac}

\def\FF{{\cal F}}

 \def\ga{\gamma}

 \def\gmn{g_{\mu\nu}}
\def\ggg{gravitational~}
\def\gr{\gtrsim}

\def\gw{gravitational wave~}
\def\gww{gravitational waves~}

\def\ha{\frac{1}{2}~}
\def\hy{hypersurface~}

\def\iiii{inertial frames~}
\def\inf{\infty}

\def\ka{\kappa}

\def\kkk{Killing field~}
\def\kkkk{Killing fields~}
\def\kv{Killing vector~}
\def\la {\lambda}
\def\lb{\label}
\def\les{\ \lesssim}
\def\lhs{left-hand side~}
\def\lll{\left(}

\def\LLL{\left[}

\def\mb{\mbox}

\def\mn{\mu\nu}

\def\na{\nabla}

\def\nn{\nonumber}
\def\nnn{\noindent}
\def\np{\newpage}

\def\om{\omega}
\def\ov {\overline}
\def\Om {\Omega}

\def\OO{{\cal O}}

\def\ppp{perturbation~}

\def\ps{\psi}
\def\ra{\rightarrow}
\def\rh{\rho}
\def\rhs{right-hand side~}
\def\rrr{\right)}

\def\RR{{\cal R}}
\def\Ra{\Rightarrow}
\def\RRR {\right]}

\def\si{\sigma}
\def\sq{\sqrt}
\def\sqg{\sqrt{-g}}

\def\sim{\simeq}
\def\sss{spacetime~}
\def\ssss{spacetimes~}

\def\sy{symmetric~}
\def\sym{symmetry~}

 \def\Si{\Sigma}

\def\SS{{\cal S}}
\def\td{\tilde}

\def\tga{\tilde\ga}

\def\te{\theta}

\def\tha{{\ts{\ha}}}

\def\tr{\tilde\rho}

\def\ts{\textstyle}
\def\tt{\tilde t}

\def\ubm{\unboldmath}

\def\vf{\varphi }

\def\vp{\varpi}

\def\vs{\vskip 0.5 cm}

\def\wrt{with respect to~}

\def\ze{\zeta}

\def\1k{\fr{1}{\ka}}

\def\2k{\fr{1}{2\ka}}

\begin{document}

\title{{\bf Gravitational waves and dragging effects}}
\author{  Ji\v r\'\i  \,\,Bi\v c\'ak$^{1,3}$\thanks{email:bicak@mbox.troja.mff.cuni.cz
} \, Joseph Katz$^{2,3} $\thanks{email:jkatz@phys.huji.ac.il
  }    \, Donald Lynden-Bell$^{3}$\thanks{email:
dlb@ast.cam.ac.uk}\\
\\ {\it $^1$Institute of Theoretical Physics,  Charles University, 180 00 Prague 8, Czech Republic} \\
\\ $^2${\it The Racah Institute of Physics, Givat Ram, 91904 Jerusalem,
Israel}
\\\\
 {\it $^3$Institute of Astronomy, Madingley Road, Cambridge CB3 0HA,
United Kingdom}
\\ }
\maketitle
\begin{abstract} 
\setlength{\baselineskip}{20pt plus2pt}
Linear and rotational dragging effects of \gww on local \iiii are studied in purely vacuum spacetimes. 
First the linear dragging caused by a simple \cy pulse is investigated. Surprisingly strong transversal effects of the pulse are exhibited. The \am in cylindrically \sy \ssss is then defined and confronted with some results in literature. In the main part, the general procedure is developed for studying weak \gww with translational but not axial symmetry which can carry angular momentum. After a suitable averaging the rotation of local \iiii due to such rotating waves can be calculated explicitly and illustrated graphically. This is done in detail in the accompanying paper. Finally, the rotational dragging is given for strong \cy waves interacting with a rotating cosmic string with a small angular momentum.

\vs
\vs
\vs

\nnn PACS numbers 04.20.Jb \,\,\, 04.30. -w

\end{abstract} 

 \np
\setlength{\baselineskip}{20pt plus2pt}
\sect{Introduction}
Bets have been won by those who prognosticated that \gww will not be positively detected in an earth-bound laboratory until 1999, 2001,..., 2008. But it would be hard to find a reader of {\it Classical and Quantum Gravity} who does not believe that eventually \gww will be detected through their effects
on test laboratory masses. From purely theoretical perspectives, from the year 2000 at least six papers appeared in this journal in which dragging effects even on gyroscopes were studied: for example, the precession of a small test gyroscope induced by a weak plane gravitational wave \cite{BD}, \cite{SBD}, the behaviour of a gyroscope located at smooth and polyhomogeneous null infinity \cite{HH}, \cite{Va}, 
the motion of spinning test particles in plane gravitational waves \cite{MS}, \cite{MTW}.
 
 However, a truly ``Machian" problem associated with \gww is not concerned with the behaviour of a gyroscope immersed directly in a gravitational wave. A collapsing, slowly rotating spherical massive shell drags into rotation \iiii inside its {\it flat} interior \cite{LB}, \cite{KLB}. In linearly perturbed Friedmann-Robertson-Walker universes, the local \iiii can be seen to be determined instantaneously via the perturbed Einstein field \eeee even from the distributions of energy, momentum and \am beyond a cosmological horizon \cite{BLK04}, \cite{BKL}. Correspondingly,  it is important to demonstrate whether energy and \am of \gww in {\it purely vacuum} \ssss can cause the local \iiii to rotate. If, indeed, this is so, it is of interest to see what such effects look like explicitly. 
  
  Thanks to progress in ``analytic gluing techniques" in geometric analysis and differential geometry over the past 25 years, and their recent applications to General Relativity, the question of the existence of the effect of the \am of \gww on the rotation of \iiii can now be answered with full rigour. 
  In our context the important {\it physical} statement following from these developments, in particular from the work of Corvino \cite{Co} and Corvino and Schoen \cite{CS}, is that \as flat initial vacuum data, i.e., data which represent pure \gww  can be deformed so that in the far region they are exactly Kerrian, i.e., stationary. Since the Kerr metric is one of the best prototypes to demonstrate the  dragging of local \iiii, there is now clear theoretical evidence available that \am due to \gww causes local \iiii to rotate \wrt the local \iiii at infinity. Moreover, as in the cosmological \ppp theory, or in the examples with rotating shells, this effect is global and instantaneous (all the construction occurs on a given spatial hypersurface). From the perspective of Mach's principle, one would prefer to demonstrate the rotation of local \iiii in a {\it closed} universe filled only with \gww since then boundary conditions (like asymptotic flateness) play no role. Such a study lies in the future. However one would also like to see the dragging effects caused by the ``rotating waves" explicitly. To construct explicit examples, we turn to the waves with a symmetry along $Oz$ in \cy polar coordinates.
 
 In the next section the metric with a translational symmetry is introduced in suitable \coo and the vacuum field \eeee in four dimensions are rewritten as Einstein's \eeee in three dimensions with a scalar field as a source, following \cite{ABS1}. A rotational symmetry is then assumed in addition, and the \eeee for standard Einstein-Rosen waves are written down, including one of their simplest explicit solution, the Bonnor/Weber-Wheeler time symmetric pulse. Section 3 is devoted to the effects this pulse exerts on test particles. Specifically, we calculate the force acting on a particle which is at rest on a cylinder with  either fixed proper circumference or a fixed proper radial distance from the axis. Interestingly, it can happen that {\it no} physical force can keep   particles at fixed circumferences   in the ``strong region" of the pulse. However, the particle can be forced to stay fixed at a given proper distance from the axis or at a given radial coordinate (defined geometrically  in   \ER spacetimes).
 The ``linear stretching" due to the \cy wave in the directions transverse to its propagation thus appears to be stronger than the ``linear dragging"   in the radial direction.
 
 In the following sections we turn to the ``rotational dragging". First, however, we analyze the \am in cylindrically \sy \ssss (Section 4). Those \ssss posses the rotational \kv which can be normalized at the axis of  \sym and used to define the total \am density, contained in a 2-dimensional cylinder of a unit length, by adopting the Komar expression. The Komar-type expression yields meaningful results even if a rotating cosmic string is present along the axis. 
We show that among \cy metrics which appeared recently in the literature the \am is non-vanishing only in those cases which represent \cy \gww interacting with a rotating cosmic string.
 
  An interplay between a strong \gw and a rotating string is of interest but a physically ``clean" explicit demonstration of dragging effects due to \gww should not involve any   matter like a string. A simple procedure for studying \gww with a net \am is presented in Section 5. We assume the waves to have the translational symmetry but not axial symmetry. Then it is convenient to make dimensional reduction after which the vacuum problem in four dimensions becomes that of a scalar field in 3-dimensional spacetime \cite{ABS1}. The scalar field satisfies the wave \eee in flat \sss provided that it is weak. We can construct ``$\vf$-dependent", rotating solutions of this flat-space wave \eee and obtain the effective \em tensor - which is of the second order - on the \rhs of the Einstein field \eeee in three dimensions. The resulting metric deviates from 3-dimensional Minkowski metric by terms of second order. 
   The rotation of \iiii on the axis cannot depend on $\vf$. We are therefore primarily interested in the behaviour with radius of the $\vf$-averaged dragging rate. We discuss the structure of all the field equations, their consistency and suitable boundary conditions within this approach. The same function in the metric which approaches a constant at spatial infinity and  represents the total energy   in the case of the \ER waves, also yields the energy of rotating waves. The explicit solution of these \eeee based on a non-axially symmetric generalization of  a weak Bonnor/Weber-Wheeler pulse, are constructed in the accompanying paper \cite{LBK07}. Therein the dragging effects of the resulting \gww on local \iiii are analyzed in detail and graphically illustrated.
In Appendix A, the Ricci tensor for the averaged rotating gravitational waves in 2+1 dimensions is given.

The rotational dragging in the spacetime of the exact (in general strong) Bonnor/Weber-Wheeler
cylindrical pulse perturbed by a rotating cosmic string with a small angular momentum is considered in the last Section 6.
 
\sect{The metric and field equations for axially rotating gravitational waves}
 
  At the core of this work and the following paper \cite{LBK07} are source-free metrics with at least one  hypersurface orthogonal Killing vector of 
  spacelike translations \bm$\ze$\ubm. In this case coordinates exist\footnote{
Greek indices $\la, \mu, \nu, \rho,\cdots$ run over the four spacetime coordinate labels $0,1,2,3$;
Latin indices $a,b,c,\cdots $ run over the time and two spatial coordinate labels $0,1,2$. The metric
$\gmn$ has signature $+---$ and $g$ is its determinant. Covariant derivatives are indicated by a $D$, partial
derivatives   by a $\di$ and covariant derivatives in a three subspace by $\na_a$. The permutation symbol in 4 dimensions is $\ep_{\mu\nu\rho\si}$ with
$\ep_{0123}=1$, $\eta_{\mu\nu\rh\si}=\sqg \ep_{\mu\nu\rho\si} $.}
  \be
  \{x^\mu\}=\{x^a, x^3=z\}~,~~~~~{\rm  ~where}~~~\mb{\bm$\ze$\ubm}=\{0, 0, 0, 1\},
  \lb{21}
  \ee
and   the metric can be written in the form  introduced by Ashtekar, Bi\v c\'ak and Schmidt \cite{ABS2}:    
  \be
ds^2=   e^{-2\psi}g_{ab}dx^adx^b-e^{2\psi} dz^2,
\lb{22}
  \ee
 $\psi$ and $g_{ab}$ are functions of $x^c$ only. The source-free Einstein's \eeee $R_{\mn}=0$ take the following interesting form in terms of the Ricci tensor $\RR_{ab}$ of the 3-space $d\si^2=g_{ab}dx^adx^b$:
 \ba
 R_{ab}&=&0~~~\Ra~~~\RR_{ab}=2\di_a\psi\di_b\psi,
 \lb{23}
 \\
 R_{33}&=&0~~~\Ra~~~g^{ab}\na_a \na_b\psi=0.
 \lb{24}
 \ea 
These \eeee can be interpreted as Einstein's  \eeee in three dimensions with a scalar field $\FF=\ps/\sq{4\pi G}$ as a source. 

 The Einstein-Rosen metrics \cite{ER} represent a particular class of   metrics of the form (\ref{22}) which admits an additional \hy orthogonal \kkk of rotations \bm$\eta$\ubm. In this case, coordinates are generally chosen   such     that {\bm$\eta$\ubm}$  =\{ 0, 0, 1, 0\} $ and   are   denoted by $x^0=t, x^1=\rho, x^2=\vf$. The metric then reads
 \be
 ds^2= e^{-2\ps}\LLL  e^{2\ga}\lll dt^2-d\rh^2\rrr -\rh^2 d\vf^2  \RRR-e^{2\ps}dz^2,
 \lb{25}
 \ee
 $\ps$ and $\ga$ are functions of $t$ and $\rh$. Notice that the radial variable $\rh$ is a  geometric object \cite{BS}. In this case the non-trivial   \eeee (\ref{23}) and (\ref{24}) can be written as follows (in the formulas, $\di_0X=\di_tX=\dot X$ and $\di_1X=\di_\rh X=  X'$ for any $X(t,\rh)$):
 \ba
 \RR_{00}&=&\ga''+\fr{1}{\rh}\ga' - \ddot\ga=2{\dot\ps}^2,
 \lb{26}\\
  \RR_{11}&=& -   \ga''+\fr{1}{\rh}\ga' + \ddot\ga =2{\ps'}^2,
   \lb{27}\\
  \RR_{01}&=&\fr{1}{\rh}\dot\ga =2\dot\ps\ps',
  \lb{28}
 \ea
  and
  \be
  \ps''+\fr{1}{\rh}\ps' - \ddot\ps=0.
  \lb{29}
  \ee
 The last \eee is that of cylindrical waves in cylindrical \coo in flat space. Given a solution $\ps$ satisfying appropriate boundary conditions, the function $\ga$ is entirely defined by the following combinations of \eeee (\ref{26}) - (\ref{28}):
  \be
  \ga'=\rh\lll  \ps'{\,^2}+\dot\ps^2  \rrr~~~{\rm and}~~~\dot\ga=2\rh\dot\ps\ps'.
  \lb{210}
  \ee
  The wave \eee (\ref{29}) guarantees that the integrability conditions of the last two \eeee are satisfied.
  
  Of particular interest are the Bonnor \cite{Bo1} and Weber-Wheeler \cite{WW} time symmetric incoming and outgoing waves which are smooth and finite everywhere at all time. They have been also  discussed in some detail  in Weber's book \cite{We}. That metric is not flat at spatial infinity. Its asymptotic behavior   received special attention in \cite{ABS2}.  The explicit  forms of $\ps$ and $\ga$ given below are taken from   Ashtekar, Bi\v c\'ak and Schmidt \cite{ABS1}; we shall write them with non-dimensional variables:    
\be
\tr=\fr{\rh}{a}~~~,~~~\tt=\fr{t}{a}~~~,~~~b=\fr{\sq{2}C}{a}.
\lb{211}
\ee  
 $C$ and $a$ are two constants of integration: $a$ is a measure of the width and $C$ a measure of the amplitude of the waves\footnote{We shall  refer to them simply as ``the width" and ``the amplitude".} \cite{We}, and \cite{WW}:
 \ba
 \ps &=&b{\Bigg \{}   \fr{1+\tr^2-\tt^2+[(1+\tr^2-\tt^2)^2+4\tt^2]^{1/2}}{(1+\tr^2-\tt^2)^2+4\tt^2}          {\Bigg \}}^{1/2}~~~,~~~\td\ps=\fr{\ps}{b},
 \lb{212}\\
 \ga&=&\fr{b^2}{4}{\Bigg \{}  1-2\tr^2\fr{(1+\tr^2-\tt^2)^2-4\tt^2}{[(1+\tr^2-\tt^2)^2+4\tt^2]^2}  - \fr{1-\tr^2+\tt^2}{[(1+\tr^2-\tt^2)^2+4\tt^2]^{1/2}} {\Bigg \}}~~~,~~~\td\ga=\fr{\ga}{b^2}\nn.
\\ \lb{213}
\ea
Figure 1 illustrates $\td\ps(t,\rh)$ and $\tga(t,\rh)$ as functions of $\tr$ for various values of $\tt$. Notice that $\td\ga\ra\ts{\ha}$ for $\rh\ra\infty$.
\begin{figure}[ht]
   \centering
      \includegraphics[width=15 cm]{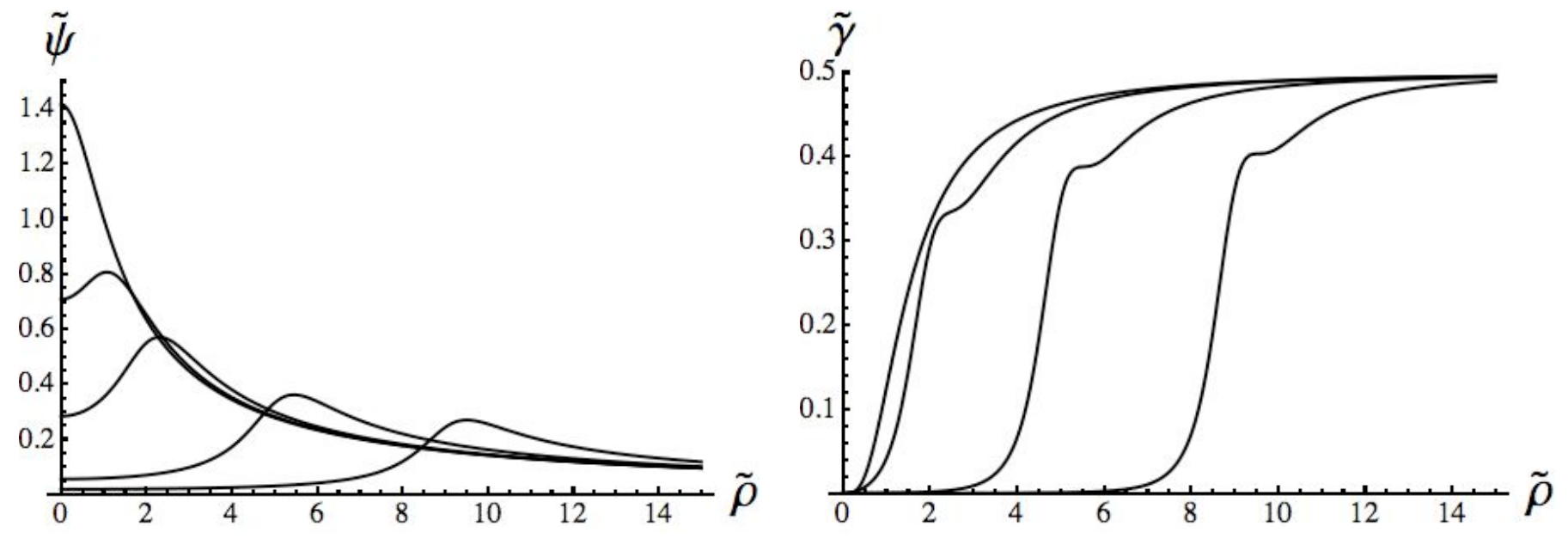} 
   \caption{\small The  plots of $\td\ps(\tt,\tr)$ for $|\tt|=0, 1, 2, 5, 9$ from the left to the right, and of $\tga(\tt,\tr)$ as functions of $\tr$ for  $|\tt|=0, 2, 5, 9$ from the  left to the right.}
   \end{figure}

\np 
\sect{The linear dragging in a Bonnor and Weber-Wheeler wave}

One way to calculate the acceleration due to gravity on earth is to calculate the force per unit mass necessary to oppose it and prevent the body from falling. To a relativist that force has the spatial components of the 4-acceleration necessary to keep the body at rest. Here we evaluate the force necessary to keep a particle in various geometrically defined positions   subject to a Bonnor/Weber-Wheeler pulse. Surprisingly we find it sometimes impossible to keep the particle on a cylinder of a fixed proper circumference length. However, it is possible to keep the particle at a  fixed proper distance from the axis or at fixed $\rh$. These worldlines are always timelike.

Let
  $v^\mu={dx^\mu}/{d\tau}$, $g_{\mn}v^\mu v^\nu=1$
represent the 4-velocity of a particle with proper time     $\tau$   which is not falling freely. Notice that the momenta of a {\it freely falling} particle in the $\vf$ or $z$ directions are constants of motion because of the global symmetries of spacetime. Thus any acceleration in the   $\vf$ or $z$ directions can only be due to non-gravitational forces. Although the wave can stretch significantly   distances between points in transverse directions, gravitational forces act only in the $\rh$ direction. The interesting cases here are thus those   of particles   forced to stay or   move  in   the $\rh$ direction. Their 4-velocity has two non-zero components $v^0$ and $v^1$. Following  (\ref{25}), 
  \be
  g_{\mn}v^\mu v^\nu=e^{2(\ga-\ps)}\LLL   (v^0)^2-(v^1)^2 \RRR=1,
\lb{32}
  \ee
 while $ v_0=e^{2(\ga-\ps)}v^0$ and $v_1= - e^{2(\ga-\ps)}v^1$.
  So \eee (\ref{32}) can also be written as
  \be
  (v_0)^2-(v_1)^2=e^{2(\ga-\ps)}.
  \lb{34}
  \ee
  The momenta per unit mass $v_2=v_3=0$. The 4-acceleration  is
  $a_\mu={Dv_\mu}/{D\tau}={dv_\mu}/{d\tau} + \tha v_\rh v_\si \di_\mu g^{\rh\si}$.
 It has two  components,
 \be
 a_0=\fr{dv_0}{d\tau}-(\dot\ga-\dot\ps)~~~{\rm and}~~~a_1=\fr{dv_1}{d\tau} - (\ga'-\ps').
 \lb{36}
 \ee
However, if the particle is forced to move along the worldline $[t(\tau),\rh(\tau)]$, so that $v_\mu=v_\mu[t(\tau),\rh(\tau)]$, we have successively
\be
\fr{dv_0}{d\tau}=\dot{v}_0v^0+{v_0}'v^1~~~{\rm and}~~~\fr{dv_1}{d\tau}=\dot{v}_1v^0+{v_1}'v^1.
\lb{37}
\ee  
  On the other hand the derivatives of (\ref{34}) give respectively
  \be
  v_0\dot{v}_0-v_1\dot{v}_1=(\dot\ga-\dot\ps)e^{\ga-\ps}~~~{\rm and}~~~  v_0 {v_0}'-v_1{v_1}'=( \ga'- \ps')e^{\ga-\ps}.
  \lb{38}
  \ee
  We may eliminate for instance $\dot{v}_0$ and ${v_1}'$ from the pair of \eeee  (\ref{37}) and  (\ref{38}). The new expressions for ${dv_0}/{d\tau}$ and ${dv_1}/{d\tau}$ may be substituted in   (\ref{36}) which becomes 
  \be
  a_0=v^1({v_0}' - {\dot v}_1)~~~{\rm and}~~~  a_1=-v^0({v_0}' - {\dot v}_1).
  \lb{39}
  \ee
  This makes it obvious that $a_\mu v^\mu=0$.
  
  The frame component of the force   $f$ per unit mass or the acceleration exerted on the particle is
  \be
  f=\sq{-g_{11}}a^1= - \sq{- g^{11}}a_1=-e^{-(\ga-\ps)}a_1.
  \lb{310}
  \ee
  The magnitude of the acceleration is
  \be
  \sq{-a_\mu a^\mu}=e^{-2(\ga-\ps)}|{v_0}' - {\dot v}_1|.
  \lb{311}
  \ee

Suppose now a   particle sits at $\vf=\vf_0$ and $z=z_0$ and its worldline is given by  
\be
F( t, \rh)=\td C~,~~~~~~  \td C~{\rm is~constant}.
\lb{312}
\ee
 If  $F'\ne0$, we may use $\td C$ instead of $\rh$ as a spatial coordinate. Differentiating (\ref{312}), we obtain
$d\td C=\dot F dt+F'd\rh$, from which follows that $d\rh=({1}/{F'})(d\td C-{\dot F}dt)$.
Substituting this $d\rh$ into (\ref{25}) we obtain
\be
ds^2= \fr{e^{2(\ga-\ps)}}{F'^2}\LLL   (F'^2-\dot{F}^2)dt^2 +2\dot{F}d\td C dt -d\td C^2  \RRR - e^{-2\ps}\rh^2d\vf^2 - e^{2\ps}dz^2.
\lb{314}
\ee
The particle in question is   at rest in $\{t, \td C, \vf, z\}$ coordinates. The proper time elapsing after an interval $dt$ is thus
$
d\tau= \left( {e^{\ga-\ps}}/{|F'|} \right) \sq{F'^2-\dot{F}^2}dt
$, and {\it the particle can only be at rest if} 
\be 
{F'}^2-{\dot F}^2>0.
\lb{316}
\ee
Otherwise $\td C=F( t, \rh)$  represents a spacelike worldline.

Assuming the inequality (\ref{316}) is satisfied, we may calculate the force per unit mass that must act on a particle to be kept at fixed $\td C$.
Since
\be
\fr{dF}{d\tau}=\dot F v^0+F'v^1=e^{-2(\ga-\ps)}(\dot F v_0-F'v_1)=0,
\lb{317}
\ee
\eee (\ref{317}) with (\ref{34}) define $v_0$ and $v_1$. One readily finds that if $v_0>0$ which is always allowed,
\be
v_0=\ep_{F'}F'H~~,~~v_1=\ep_{F'}\dot FH~~{\rm with}~~\ep_{F'}=sign(F') ~~, ~~H=\fr{e^{\ga-\ps}}{\sq{{F'}^2-{\dot F}^2}},
\lb{318}
\ee 
and the force per unit mass is given by  (\ref{310})  or
\be
f=e^{\ps-\ga}v^0({v_0}' - {\dot v}_1).
\lb{319}
\ee

Now the particle  will stay on a circumference  of length $l$ if
$F=\rh e^{-\ps}={l}/{2\pi}$, or in dimensonless form, $L=\tr e^{-\ps}={l}/{2\pi a}$.
Then, $\di_{\tr} L =(1-\rh\ps')e^{-\ps}$ and $ \di_{\tt} L =-\rh\dot\ps e^{-\ps}$,
and condition (\ref{316}) amounts to having
$(\di_{\tr} L)^2-(\di_{\tt} L)^2>0$, or
\be
U(\tt, \tr)=(1-\rh\ps')^2-(\rh\dot\ps)^2>0.
\lb{323}
\ee
We are inter  $U(\tt,\tr)$ as a function of   $L(\tt,\tr)$.
 We consider first an amplitude $b=1$. We shall see later what happens when $b$ is not equal to $1$.  

$L(\tt,\tr=0)=0$ and  $L(\tt,\tr=\inf)=\inf$. $L$ increases monotonically for any $|\tt|$   less than about 6.  For $|\tt|\gr6$, $L$ decreases in some interval say $\{\tr_1, \tr_2\}$ which depends on $|\tt|$ and is always around $\tr=|\tt|$; see Figure 2.

\begin{figure}[ht]
   \centering
   \includegraphics[width=16 cm]{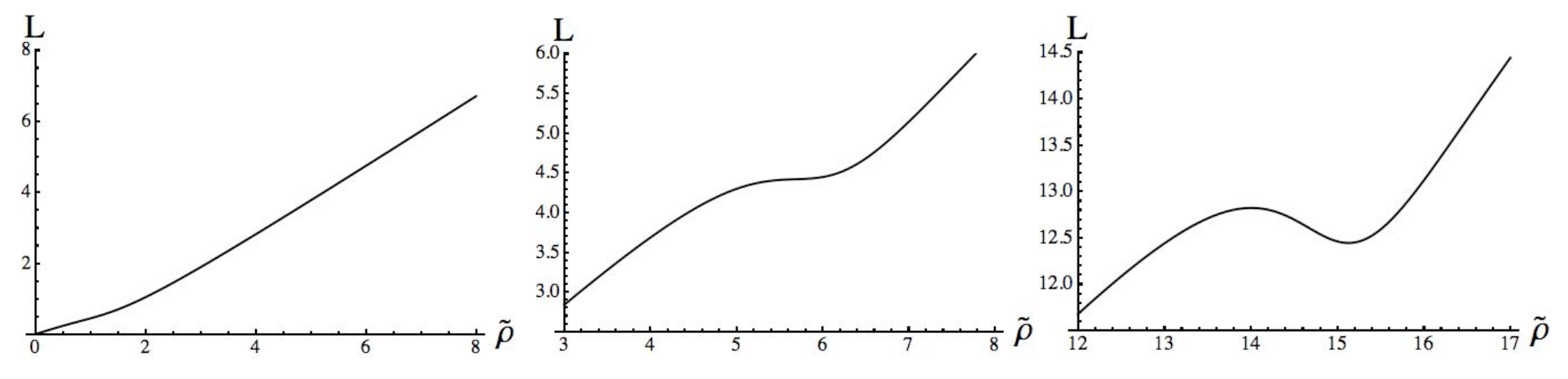} 
   \caption{\small  From left to right $L$ as a function of $\tr$ for $|\tt|=1, 6,$ and $15$. }
   \end{figure}
 
 Now consider $U$.   $U(\tt,\tr=0)=1$ and $U(\tt,\tr=\inf)=1$.  Thus $U$ is positive for small and big  $L$'s. What happens for intermediate $L$'s? 
 \begin{figure}[ht]
 \centering
 \includegraphics[width=12 cm]{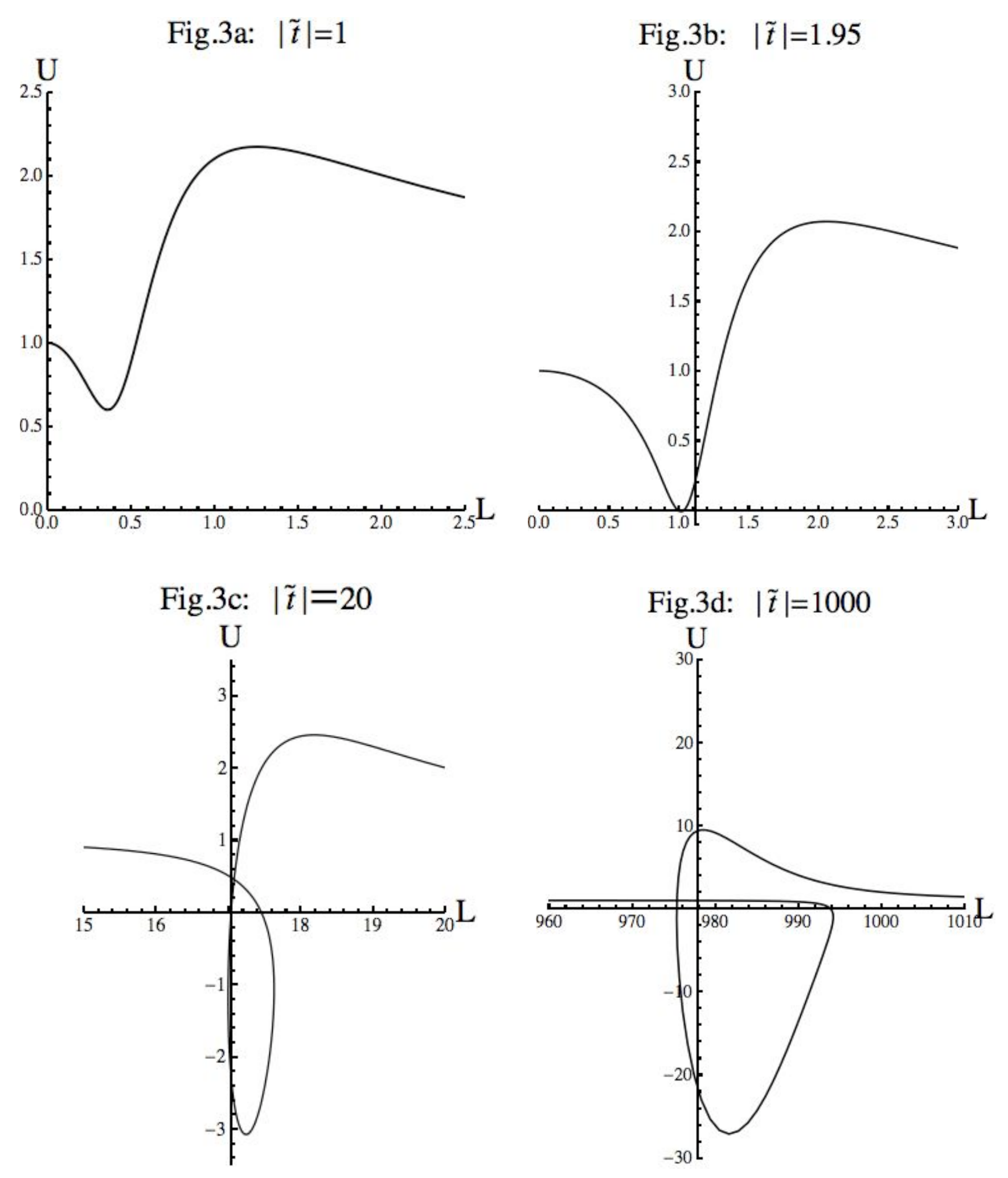} 
   \caption{\small $U$ as a function of   $L$. A particle can stay at a given $L$ only if $U>0$. See text for details.}
   \end{figure}
Consider  $U$  as a function of $L$ for various values of the dimensonless  time $\tt$. For $ |\tt|\les 2$, $U>0$. $U$ varies monotonically when $\rh$ nears $0$ and $\infty$. A typical example is given in Figure 3a where $|\tt|=1$. Notice that in the graphs of Figure 3 the $U$ axis cuts the $L$ axis where $|\tt|=\tr$. This is done for    reasons appearing later.  Figure 3b shows how $U(L)$ varies when $|\tt|\sim1.95$; this is a limiting case.  It becomes  impossible to keep a particle on some of the  cylinders at $|\tt|\gr1.95$. For instance  for $|\tt|=6$, when $L$ has a horizontal inflexion, $U<0$ for  $4.22\les L\les 4.56$ corresponding to an interval $4.83\les\tr\les6.32$. 
For $|\tt|\gr6$,   $U(L)$ has a  loop because $L$ does no more increase monotonically; see for instance Figure 3c in which $|\tt|=20$. At that time it becomes impossible to keep a particle at   $17.05\les L \les17.4$ corresponding to $18.1\les\tr\les20$.  The loop persists at any later $(\tt\gr6)$  or earlier $(\tt\les-6)$ times and the knot is always where $U>0$ while $U$ is always negative for $|\tt|=\tr$. See for instance Figure 3d  for $|\tt|=1000$.

   What if the amplitude $b$  is different from $1$? For $b\ne1$ the phenomena is essentially the same. Loops appear at bigger $L$'s for   $b<1$ and at small $L$'s for  $b>1$. It may be noted that the $U$ axis, which cuts the $L$ axis where $|\tt|=\tr$ also cuts the loop in the region where $U<0$. It is therefore interesting to evaluate the upper bound on $b$ resulting from  the condition (\ref{323}) for $|\tt|=\tr$. In this way we find that $U$ is always negative   when $|\tt|=\tr$ if 
 \be
 b>b_M=\fr{\sq{(1+4\tr^2)\sq{1+(1+4\tr^2)}}}{2\tr^2}.
 \lb{324}
 \ee
Thus, the phenomena described for $b=1$ exists at any amplitude however small since for $\tr\ra\inf$, $b_M\ra2^{5/4}/\tr^{1/2}\ra0$.  

If a particle is kept at a fixed proper distance $\la=$const from the axis, then
condition (\ref{316}) becomes
\be
- e^{\ga-\ps}< \di_t\lll\int_0^{\rh} e^{\ga-\ps} d\rh\rrr<e^{\ga-\ps}.
\lb{327}
\ee
The situation is even clearer, and much easier to manage numerically,  if   the particle is kept at fixed\footnote{See \cite{BS} where it is shown that $\rh$ is a geometrically well defined quantity.} $\rh$. Since $\rh$ is a spatial \co  there is   manifestly no condition like (\ref{316}). In this case,
$v^1=0$, $v_0=e^{\ga-\ps}$,
and the  dimensionless force is given  by
$\td f=af=-\di_{\tr}e^{-(\ga-\ps)}$.
In real value, $f=\fr{c^2}{a}\td f$.
The particle can always be kept at fixed radial coordinate $\rho$.

As an example, let $\td f\sim 0.3$, a characteristic value for $|\tt|\gr 20$ and $b=1$. Let $f$ be barely bearable, say $f=5g\sim 5\cdot 10^3$ cm sec$^{-2}$. Then,
$a=\fr{c^2\td f}{f}\sim 3\cdot 10^3 \rm AU$.
The amplitude $\sq{2}C=ab$ is of the same order as $a$. The characteristic time $t_C$ during which the acceleration must be withstood is $t_C\sim \fr{a}{c}\sim 2\cd 10^6 {\rm sec}\sim {\rm 3 weeks}$.
\sect{Angular momentum of cylindrically symmetric \ssss}

Here we consider \ssss with two spacelike \kkkk \bm$\ze~$\ubm and \bm$\eta$\ubm.   \bm$\ze~$\ubm  is  in general \hy orthogonal but rotational \bm$\eta~$\ubm is in general not, so that the two-dimensional isometry group is not orthogonally transitive \cite{ES}. The \sss may admit an additional Killing vector \bm$\xi~$\ubm which is timelike and may thus represent a stationary field of a rotating massive cylinder or an infinite straight rotating cosmic string. 

Although in the main part, Section 5,  we require the axis of symmetry to be regular, in Section 6 we also discuss some aspects of cylindrical \ssss containing an infinitely thin string in the ``wire approximation". It is associated with specific conical singularities along the axis which, though being ``quasi-regular",  do not strictly speaking, belong to the spacetime. An extensive literature on cylindrical waves interacting nonlinearly with cosmic strings is available\footnote{See the review \cite{An} and references below.}. The 
wave solutions {\it in vacuo} are considered by some authors as too restrictive. The reason is a generalization of Papapetrou's theorem from stationary axisymmetric vacuum \ssss to the \cy case: if the \sss contains at least part of a regular axis of \cy symmetry, the isometry group must be orthogonally transitive, so Killing orbits admit orthogonal surfaces \cite{BCM}, \cite{CSV}, \cite{Me}. The orthogonal transitivity thus excludes the possibility of a global rotation. In other words {\it in vacuo} there can be no `rotating \cy waves'. With a material source present as  in the case of the rigidly rotating dust cylinder \cite{Bo2}, for example, the \sss can, of course, be regular everywhere with a non-vanishing \am per unit length. Bondi \cite{Bondi} studied general changes in time of such systems which can lead to radiation.
As he noticed, the conservation of \am occurs even if gravitational waves are emitted by the cylinder since the \cy symmetry of the waves precludes their carrying angular momentum.

In all the cases mentioned, the \ssss posses the rotational \kv \bm$\eta~$\ubm with closed orbits. This can be normalized at the axis of symmetry, with corresponding modification in case of a string present along the axis. When the rotational \kv is available we can define a total \am density contained inside a 2-dimensional cylinder $\CC$ of `unit coordinate height', generated by orbits of the isometry group, by adapting  the Komar \cite{Ko} expression\footnote{For using   Komar's  expression in case of the angular momentum, see \cite{Wi} and  \cite{Wa} p.296.} to   \cy symmetry:
\be
J(\CC)=\2k\int_\CC D^{[\mu}\hat\eta^{\nu]}d\Si_{\mn},~~~{\rm where}~~~d\Si_{\mn}=\ts{\ha}\ep_{\mu\nu\rh\si}dx^\rh \wedge dx^\si ~~,~~\hat\eta^{\nu}=\sq{-g}\eta^\mu~~,~~\ka=\fr{8\pi G}{c^2}
\lb{41}
\ee
Since
\be
\hat\jmath^\mu=\2k \di_\nu  \lll  D^{[\mu}\hat\eta^{\nu]} \rrr
\lb{42}
\ee
is conserved, $\di_\mu \hat\jmath^\mu=0$, in vacuo the integral (\ref{41}) is independent of the choice of $\CC$; the integral over `bottom' and `top' of the solid cylinder formed by two cylinders $\CC$ and $\CC'$ vanishes due to   \cy symmetry. This corresponds to the well known fact that the Komar energy, defined as (\ref{41}) with \bm$\eta~$\ubm being replaced by the timelike \kkk \bm$\xi~$\ubm in stationary spacetimes, is independent of the choice of the closed 2-surface as it is moved continuously through a vacuum region \cite{Wa}. Hence, we can define the total \am   per unit length in the  \bm$\ze~$\ubm direction by
\be
J=\2k\int_{\CC_\infty}D^{[\mu}\hat\eta^{\nu]}d\Si_{\mn},
\lb{43}
\ee
where $\CC_\infty$ denotes the cylinder  of unit length at infinity (there the proper and coordinate lengths along the \bm$\ze~$\ubm direction coincide).

Let us express the \am (\ref{43}) (``density" will often be omitted) explicitly for some specific metrics. In all cases the coordinates adapted to the symmetry are those defined in Section 2 below equation (\ref{24}). The 2-surface element 
\be
d\Si_{01}=-d\Si_{10}=d\vf dz.
\lb{44}
\ee 
Hence the \am (\ref{43}) reads
\be
J=\1k\int_{\CC_\infty}D^{[0}\hat\eta^{1]}d\vf dz.
\lb{45}
\ee
The integrand does not depend on $\vf$ and  $z$.  Integrating we thus get
\footnote{Let us remark that with $z$ replaced by $\te\in [0,\pi]$ and $\{t,\rh ,\te ,\vf\}$ denoting the Boyer-Lindquist coordinates in the Kerr metric (e.g. \cite{Wa}) with mass $M$ and \am per unit mass $a$, the expression (\ref{46}) yields indeed $J=Ma$.}
\be
J=  \fr{2\pi}{\ka} \LLL D^{[0}\hat\eta^{1]}\RRR_{\rh\ra\infty}.
\lb{46}
\ee

Mashhoon {\it et al} \cite{MCQ} considered a class of ``rotating gravitational waves" which should represent ``radiation that propagates outward or inward and at the same time has non-trivial azimuthal motion". Their metrics satisfying vacuum field \eeee have the form
\be
 ds^2=e^{2\ga-2\ps}(dt^2-d\rh^2)-W^2e^{-2\ps}(d\vf+\om dt)^2 - e^{2\ps}dz^2;
\lb{47}
\ee
the functions $\ga, \ps, W$ and $\om$ depend on $t$ and $\rh$ only (in   \cite{MCQ} $\rh=R$ and $W=\mu$). If $\om=0$, the field \eeee imply that $W$ can be chosen equal to $\rh$, and the metrics (\ref{47}) become the well known metrics of Einstein-Rosen waves (e.g. \cite{ES}, \cite{ABS2}). Calculation of the \am (\ref{46}) yields 
\be
J=\fr{\pi}{\ka}  W^3e^{-2\ga}\om' \Big|_{\rh\ra\inf}.
\lb{48}
\ee

In Section 6 we shall especially need the \am for small $\om$. One of the field \eeee (cf. \cite {MCQ} or the Appendix    where the Ricci tensor in the reduced 3-dimensional \sss is given) then implies $W=\rh$ (like in the case of Einstein-Rosen waves), and $\om=\vp_\inf/\rh^2+\OO(\rh^{-3})$. The \am then becomes
\be
J= \fr{\pi}{\ka}\rh^3e^{-2\ga} \om'\Big |_\inf = - \fr{2\pi}{\ka}e^{- 2\ga_\inf}\vp_\inf~~~{\rm or}~~~\vp_\inf= - \fr{\ka}{2\pi}Je^{2\ga_\inf}.
\lb{49}
\ee

Next, consider the case of infinitely thin cosmic strings.   When they posses an \am $J$, the corresponding metric can be written as  (e.g. \cite{DJH}, \cite{GL}) 
\be
ds^2=(dt+\fr{\ka}{2\pi}Jd\vf)^2-(\ov\rh)^{-\ka\mu/\pi}(d\ov\rh^2+\ov\rh^2d\vf^2)-dz^2,
\lb{410}
\ee
$\mu$  being the mass per unit length. After putting $(\ov\rh)^{-4\mu}d\ov\rh=d\rh$, the metric 
can be rewritten as
\be
ds^2=(dt+\fr{\ka}{2\pi}Jd\vf)^2-d\rh^2-(1-\fr{\ka\mu}{2\pi})^2\rh^2d\vf^2-dz^2.
\lb{411}
\ee
 In both metrics $\vf\in<0,2\pi)$\footnote{The transformation $\ov\vf=(1-\fr{\ka\mu}{2\pi})\vf$ accompanied with the change of time $\ov t=t+\fr{\ka }{2\pi}J\vf$ leads to the flat metric with the range $\ov\vf\in[0,(1-\fr{\ka\mu}{2\pi}) 2\pi)$, corresponding to conical geometry and the time coordinate `jumping' by $\ka J$ when the string is circumrotated.}. 
 Comparing (\ref{410}) and (\ref{411}) with the metric (\ref{47}) and the \am (\ref{49}) for small $\om$, resp. $J$, it is seen immediately that the Komar expression gives indeed $J$  appearing in the metric.
 
 Various authors \cite{Xa1}, \cite{Xa2}, \cite{ET}, \cite{DGN} constructed new families of exact solutions of the vacuum Einstein (and Einstein-Maxwell) \eeee for cylindrically \sy non-stationary  \ssss which they interpreted as gravitational (and electromagnetic) \cy waves interacting with a rotating cosmic string. In their solutions, however, the cross-term $dtd\vf$ is missing.
Applying our expression (\ref{46}) for the \am to their metrics, we get $J=0$. 
More recently, Manojlovi\' c and Marug\' an \cite{MM}
 considered thoroughly \cy waves interacting with spinning cosmic strings within the Hamiltonian formalism. In their work the non-diagonal terms proportional to $dt d\vf$ are admitted -- their value of the angular momentum coincides with that following from (\ref{49}). 
Let us also state that our expression (\ref{46}), taken at fixed $\rh$, when applied to Bondi's \cy metric (\eee (1) in \cite{Bondi}) given in the Weyl-Papapetrou form, yields precisely the Bondi expression for the angular momentum.
 
 To the end of this section let us mention one  plausible feature of the \am defined above. Imagine we transform the metric (\ref{47}) into rotating axes by introducing $\ov\vf=\vf+\Om t$, in which $\Om$ may depend on time. Then $d\ov\vf=d\vf+\Om(t)^{\bf{.}}dt$, but $ \ov\om'= \om'$ so that the total \am (\ref{49}) remains unchanged. It thus characterizes an intrinsic property of the field.
 
\sect{Rotating waves in the symmetry reduced general relativity}

To construct rotating \ggg waves with a non-vanishing \am we can keep the translational symmetry but we have to drop the assumption of axial symmetry. In general we would have to solve the coupled system of \eeee (\ref{23}) and (\ref{24})
with $\ps$ and $g_{ab}$ depending on all three coordinates $x^0=t, x^1=\rh, x^2=\vf$. In 3-dimensional \ssss the Riemann tensor 
\be
\RR_{abcd}=2\LLL  \lll  \RR_{a[c}-\ts{\fr{1}{4}}g_{a[c}\RR  \rrr g_{d]b} -  \lll  \RR_{b[c}-\ts{\fr{1}{4}}g_{b[c} \RR \rrr g_{d]a}\RRR.
\lb{61}
\ee
The Ricci tensor can be determined directly from the field $\ps$ by \eeee (\ref{23}).

Clearly to tackle such a problem is a formidable task. However, the above form of Einstein's \eeee suggests naturally the following approximation procedure. Assume the field $\ps$ and its derivatives to be small, $\ps=\ep\Psi(x^a)$, $\ep$
is a small dimensionless parameter.  Then the field \eeee (\ref{23}) show that $\RR_{ab}\thicksim \OO(\ep^2)$ and, according to (\ref{61}), the Riemann tensor can be calculated in terms of $\di_a\ps$ up to   $\OO(\ep^2)$. We may thus write the 3-dimensional metric in the form
\be
g_{ab}=\eta_{ab}+\ep^2\ga_{ab}(x^c)~~~{\rm or}~~~g^{ab}=\eta^{ab}-\ep^2\ga^{ab}(x^c).
\lb{62}
\ee
  Therefore we can construct a {\it rotating} (``$\vf$-dependent") solution of the wave \eee in {\it flat space} in \cy \coo $\{ t,\rh,\vf \}$ and still satisfy the field \eeee 
 (\ref{24}) 
  in terms $\thicksim \OO(\ep^2)$. 
  
  Next, we can write down the Ricci tensor on the \lhs of the \eeee (\ref{23})
  for a general $\vf$-dependent metric, $g_{ab}=g_{ab}(t,\rh,\vf)$, and solve the \eeee in terms  which are  $\thicksim \OO(\ep^2)$. However, we are primarily interested in the rotation of \iiii at the axis where  $\om$ is not $\vf$-dependent.   Therefore we calculate the average of $\om$ over $\vf$ at each radius. To find this we consider the average over  $\vf$ of the \rhs of (\ref{23}) and obtain an effective matter source that is axially symmetric of the form 
    \be
  \SS_{ab}=<\di_a\ps\di_b\ps>=\int_0^{2\pi}\di_a\ps\di_b\ps d\vf.
  \lb{63}
  \ee
  Since the source becomes then   axisymmetric, we consider the metric $g_{ab}$, resp. $\ga_{ab}$ axisymmetric as well. However, the \kv\bm$\eta~$\ubm is not in general \hy orthogonal, and we cannot use the vacuum \eeee to simplify the choice of coordinates. A sufficiently general axi\sy 3-metric has the form
  \be
  d\si^2=e^{2\ga}\lll  dt^2 - d\rh^2   \rrr -W^2\lll  d\vf - \om dt   \rrr^2,
  \lb{64}
  \ee
where $\ga, W$ and $\om$  
   are functions of $t$ and $\rh$. Notice that this is precisely the 4-metric (\ref{47})
   reduced to the 3 dimensions and rescaled by the norm of the Killing vector \bm$\ze$\ubm. The metric (\ref{64}) satisfies the relations $g_{00}= - g_{11}, \, g_{01}=g_{12}=0$, which can be achieved by a suitable choice of coordinates. The Ricci tensor components for the exact metric (\ref{64}) are listed in Appendix A. Within our approximation scheme it is sufficient to consider only terms $\thicksim \OO(\ep^2)$. Since for $\ep=0$ the metric (\ref{64}) becomes flat, we can write 
   \be
   W(t,\rh)=\rh+\ep^2w(t,\rh),
   \lb{65}
   \ee
   where $w \thicksim\OO(\rh^2)$ at $\rh\ra0$, but it will be convenient to work with $W$. Henceforth, $\ep^2$ is absorbed in the corresponding symbols, so it is assumed that $\ps$ is of the order $\OO(\ep)$, and $\ga,w$ and $\om$ and their derivatives are all of  $\OO(\ep^2)$; their products can be neglected. From the expressions for $\RR_{ab}$ in   Appendix  we get, to order $\OO(\ep^2)$, the following set of equations:
   \ba
   \RR_{00}&=& - \ddot\ga+\ga'' +\fr{1}{\rh}  \ga'-\fr{1}{\rh} \ddot W~~,~~ \RR_{01}=\fr{1}{\rh}  \dot\ga-\fr{1}{\rh}{\dot   W}'~~,~~   \RR_{02}=\fr{1}{2\rh}(\rh^3\om')',
   \lb{66}\\
   \RR_{11}&=&\ddot\ga - \ga'' +\fr{1}{\rh}  \ga'-\fr{1}{\rh} W''~~,~~ \RR_{12}=\tha\rh^2{\dot\om}'~~,~~ \RR_{22}=\rh(\ddot W - W'').
    \lb{67}
   \ea
  The wave \eee (\ref{24})
for the field $\ps(t,\rh,\vf)$ becomes, in our approximation, the flat space wave \eee in \cy coordinates:
\be
\ddot\ps - \ps''-\fr{1}{\rh}\ps'-\fr{1}{\rh^2}\di_\vf^2\ps=0.
\lb{68}
\ee  
For $W=\rh, ~\om=0$ we recover the Einstein-Rosen waves. In their case one can determine $\ga(t,\rh)$ by a quadrature. Now the \eeee
\be
\RR_{ab}=2<\di_a\ps\di_b\ps>=2\SS_{ab},
\lb{69}
\ee  
 with $\RR_{ab}$ given by (\ref{66})   and (\ref{67})
  should determine the functions $\ga, W$ and $\om$.
  
  It is well-known that the \ggg field in the 3-dimensional spacetime has no dynamical degrees of freedom; these are all contained in the matter fields, in our case in the field $\ps$. The entire curvature (\ref{61}) is determined by the local distribution of $\ps$. Outside matter the spacetime is flat. In the canonical gravity language, the 2-dimensional metric and the conjugate momenta have six independent components, three of which can be annulled by the choice of a spacelike \hy (time) and two suitable \coo on the hypersurface. In addition, there are three constrained field \eeee (see e.g. \cite{GAK} for an interesting exposition of Einstein's theory in a 3-dimensional spacetime).
  
  We have chosen suitable \coo
  in the metric (\ref{64}) already. Hence, we have to look at the constraints. One is evident. The \eee
  \be
  \RR_{02}=\fr{1}{2\rh}\lll \rh^3\om'   \rrr'=2<\dot\ps\di_\vf\ps>,
  \lb{610}
  \ee
   following from (\ref{66}), gives with suitable boundary conditions the `angular velocity' $\om(t,\rh)$ in terms of the averaged `angular momentum' component $\SS_{02}$ of rotating $\ps$-waves. This equation will be analyzed in depth in the accompanying paper \cite{LBK07}. In the explicit construction we shall demonstrate how rotating \ggg waves rotate local \iiii in a `global sense', like rotating `ordinary' matter does. Alternatively, the integration function of time is fixed by the \eee $\RR_{12}=2\SS_{12}$, see (\ref{67}). The \eee for $\om$ still leaves $\om$ undetermined up to an arbitrary function of time $\om_0(t)$. This arbitrariness, however, corresponds just to going over to rotating axes (cf. the end of Section 4). Another constraint equation, $\RR_{01}=2\SS_{01}$,  see (\ref{66}), can be written as 
 \be
\fr{1}{\rh}\dot\FF= 2\SS_{01}~~~{\rm where}~~~\FF=\ga-W'.
\lb{611}
 \ee
 This becomes (\ref{28}) for Einstein-Rosen waves ($W=\rh$). 
 
 There is still a constraint equation, the one corresponding to the Hamiltonian constraint in canonical gravity. Its \lhs is given by $\RR_{00}-\ts{\fr{1}{2}}g_{00}\RR$ which, in our approximation, is equal to $\ts{\fr{1}{2}}(\RR_{00}+\RR_{11}+\rh^{-2}\RR_{22})$. Substituting for the Ricci tensor from (\ref{66}) and (\ref{67}), we obtain the constraint in the form
 \be
 \fr{1}{\rh}\FF'=\fr{1}{\rh}(\ga - W')'=\SS_{00}+\SS_{11}+\fr{1}{\rh^2}\SS_{22}.
 \lb{612}
 \ee

Before we indicate how to solve the field equations, we investigate the consistency of (\ref{610}) and (\ref{611}). This should generalize the condition $(\dot\ga)'=(\ga')^{\bf{ .}}$ for the Einstein-Rosen waves when $\dot\ga$ is expressed from 
(\ref{28}), whereas $\ga'$ from the sum of (\ref{26}) and (\ref{27}). There the consistency is guaranteed by the wave \eee (\ref{29}) for the source $\ps$. Now $\ps$
satisfies the flat-space wave equation (\ref{68}) in \cy \coo with $\di_\vf\ps\ne0$. As a consequence of this \eee the effective \em tensor
 \be
T_{ab}=\di_a\ps\di_b\ps - \ts{\fr{1}{2}} \Bg_{ab}\Bg^{cd}\di_c\ps\di_d\ps~~~,~~~\Bg_{ab}=(1,-1,-\rh^2),
\lb{613}
\ee
 satisfies the conservation law 
 \be
 \na_bT_a^b=\fr{1}{\sq{-\Bg}}\di_b\lll  \sq{-\Bg} T_a^b \rrr -\ts{\fr{1}{2}} T^{cd}\di_b\Bg_{cd}=0.
 \lb{614}
 \ee
 Denoting by $s_{ab}=\di_a\ps\di_b\ps$ {\it without} the averaging so that $<s_{ab}>=\SS_{ab}$, we easily get the relations
 \be
 T_{ab}=s_{ab}-\ts{\fr{1}{2}}  sg_{ab}~~~\Ra~~~s_{ab}=T_{ab}-  Tg_{ab}~~~{\rm where}~~~  s=-2T=\dot\ps^2-{\ps'}^2-\fr{1}{\rh^2}(\di_\vf\ps)^2.
 \lb{615}
 \ee
  The conservation law (\ref{614}) written in terms of $s_{ab}$ reads explicitly as follows
  \ba
  \dot s_{00} - \fr{1}{\rh}s_{01}-s'_{01}-\fr{1}{\rh^2}\di_\vf s_{02} - \ts{\fr{1}{2}} \dot s&=&0, 
  \lb{616}\\
    \dot s_{01} - \fr{1}{\rh}s_{11}-s'_{11}-\fr{1}{\rh^2}\di_\vf s_{12} +\fr{1}{\rh^3}s_{22}- \ts{\fr{1}{2}}   s'&=&0,
\lb{617}\\
  \dot s_{02} - \fr{1}{\rh}s_{12}-s'_{12}-\fr{1}{\rh^2}\di_\vf s_{22} - \ts{\fr{1}{2}} \di_\vf s&=&0.
\lb{618}
  \ea
 The averaging over $\vf$ makes  the $\di_\vf s_{ab}$-terms to vanish so that $\SS_{ab}$ satisfies the \eeee
 \ba
   \dot \SS_{00} - \fr{1}{\rh}\SS_{01}-\SS'_{01} - \ts{\fr{1}{2}} \dot \SS&=&0~~,~~
 \dot \SS_{01} - \fr{1}{\rh}\SS_{11}-S'_{11} +\fr{1}{\rh^3}\SS_{22}- \ts{\fr{1}{2}}   \SS'=0,
\lb{619}\\
\dot \SS_{01} - \fr{1}{\rh}\SS_{12}-\SS'_{12}  &=&0~~{\rm with} ~~\SS=\SS_{00}-\SS_{11}-\fr{1}{\rh^2}\SS_{22}.
\lb{620}
 \ea
   The compatibility of   (\ref{611})  and (\ref{612}) requires
 \be
( \dot\FF)'=2(\rh\SS_{01})'=(\FF')^{.}=\rh(  \dot\SS_{00}+\dot\SS_{11}+\fr{1}{\rh^2}\dot\SS_{22}).
\lb{621}
 \ee
 This is indeed satisfied as a consequence of  (\ref{619}).
  
 What boundary conditions do we impose? As emphasized above, all the dynamics is governed by the field $\ps(t,\rh,\vf)$ satisfying the wave \eee (\ref{68}). We consider decaying solutions at spatial infinity which guarantee that the 3-dimensional \sss is asymptotically flat although the metric at spatial infinity has a ``conical singularity" which is the measure of the total energy of the scalar field $\ps$ computed using the Minkowski metric. This will generalize naturally the case of the Einstein-Rosen waves (see \cite{ABS2} for the details). Hence we assume the solutions of (\ref{68}) to have the expansion on each Cauchy surface $t$=const of the form
 \be
 \ps(t,\rh,\vf)=\fr{1}{\rh^{1/2}}\LLL   f_0(t,\vf)+\sum_ {k=1}^{\inf}\fr{f_k(t,\vf)}{\rh^k}           \RRR.
 \lb{622}
 \ee
The solutions of the wave \eee in 2+1 dimensions well-behaved at infinity involve the Bessel functions which decay as $\rh^{-1/2}$. Our explicit solutions constructed in the accompanying paper \cite{LBK07} will all admit the expansion of the form (\ref{622}). 
 
 Since the \sss should be \as flat at infinity the function $W$ must behave as follows:
 \be
 W(t,\rh)=\rh+w_0(t)+\sum_ {k=1}^{\inf}\fr{w_k(t)}{\rh^k}.
 \lb{623}
 \ee 
 However, regarding the field \eee (\ref{67}),
 \be
 \RR_{22}=\rh(\ddot W - W'')=2\SS_{22},
 \lb{624}
 \ee
 we see that $w_0(t)=w_1(t)=0$ since $\SS_{22}/\rh \thicksim \rh^{-2}$ as a consequence of (\ref{622}). Hence, we may write 
 \be
 W(t,\rh)=\rh+\sum_ {k=2}^{\inf}\fr{w_k(t)}{\rh^k}.
 \lb{625}
 \ee
 A similar argument based on the field \eee $\RR_{02}=2\SS_{02}$ implies the following behavior of the `dragging factor' $\om$:
 \be
 \om(t,\rh)=\fr{\vp_\inf(t)}{\rh^2}+\sum_ {k=1}^{\inf}\fr{\pi_k(t)}{\rh^{k+2}}.
 \lb{626}
 \ee
Since $\SS_{12} \thicksim \rh^{-2}$, the \eee $\RR_{12}=2\SS_{12}$ with $\RR_{12}$ given in (\ref{67}) implies 
\be
\vp_\inf(t)={\rm const.}
\lb{627}
\ee
 This constant is directly related to the total \am in the waves [see \eee  (\ref{49})].
 
 Now in order to guarantee a regular axis, without a conical singularity present, we require the following conditions at $\rh=0$:
 \be
 W(t,0)=\rh~~,~~W'(t,0)=1~~,~~\om(t,0)=\om_0(t)~~,~~\om'(t,0)=0~~,~~\ga(t,0)=0.
 \lb{628}
 \ee
 The only missing boundary condition is that on $\ga$ at infinity. To find it, we return back to the ``Hamiltonian" constraint, multiply by $\rh$
 and integrate to obtain
 \be
 \ga(t,\rh)- \ga(t,0)-W'(t,\rh)+W'(t,0)=\int_0^\rh\ov\rh\lll  \SS_{00}+\SS_{11}+\fr{1}{{\ov\rh}^2}  \SS_{22} \rrr d\ov\rh+\la(t),
 \lb{629}
 \ee 
 where $\la(t)$
is an arbitrary function of time, and $\ga(t,0)=0, W'(t,0)=1$ by the boundary conditions   (\ref{628}). Regarding the behavior of $W(t,\rh)$
in accordance with (\ref{625}) and realizing that $\SS_{0 1}=<\dot\ps\ps'> \thicksim \rh^{-2}$, we can deduce from the constraint  (\ref{611}) that $\la(t)$ must be a constant. Since, in addition, we have to satisfy the boundary condition $\ga(t,0)=0$, this constant has to vanish and the integrated constraint \eee (\ref{629}) becomes
\be
\ga(t,\rh)-W'(t,\rh)+1=\int_0^\rh\ov\rh\lll  \SS_{00}+\SS_{11}+\fr{1}{{\ov\rh}^2} \SS_{22}  \rrr d\ov\rh.
\lb{630}
\ee
 At $\rh=0, W'(t,0)=1$ and, indeed, $\ga(t,0)=0$. Now for $\rh\ra\inf, W'(t,\inf)=1+\OO(\rh^{-3})$, [cf. (\ref{625})], so putting $\ga(t,\inf)\eq \ga_\inf$, we get
 \be
 \ga_\inf = \int_0^\inf \lll  \SS_{00}+\SS_{11}+\fr{1}{{\rh}^2} \SS_{22}  \rrr\rh d\rh= \int_0^\inf \lll  <{\dot\ps}^2>+<{\ps'}^2> + \fr{<(\di_\vf\ps)^2>  }{{\rh}^2}\rrr \rh d\rh.
 \lb{631}
 \ee
 Therefore {\it the function $\ga$ at spatial infinity approaches a constant which is equal   to the total energy of the scalar field $\ps$ computed using the Minkowski metric}. In the case of the Einstein-Rosen waves, the term  $\rh^{-2}<(\di_\vf\ps)^2> $ vanishes, and we recover the result (2.35) in \cite{ABS2}. Although $\ga_\inf$ is energy for weak fields, the physical Hamiltonian turns out to be a non-polynomial    function of  $\ga_\inf$ for in general strong fields \cite{AV}, \cite{ABS2}.
 
 In order to find the complete solution of the averaged Einstein \eeee up to order 
$ \OO(\ep^2)$ we have to solve the \eee  (\ref{624}) with (\ref{628}), and with the \rhs given in terms of $\ps$. The function $\ga$
 can then be determined from (\ref{630}). Alternatively, we can take the combination $\ts{\fr{1}{2}}(-\RR_{00}+\RR_{11}-\rh^{-1}\RR_{22})$ and obtain an \eee for $\ga$, similar to (\ref{624}) for $W$, in the form 
 \be
 \ddot\ga - \ga''=\SS_{11}-\SS_{00}+\fr{1}{\rh^2}\SS_{22}.
 \lb{632}
 \ee
 Both \eeee (\ref{624}) and (\ref{632}) are one dimensional wave \eeee with given \rhs   and boundary conditions $\ga=\ga_\inf,\, W=1$ at $\rh\ra\inf$ and $\ga=0,\, W=1$ at $\rh=0$. Both \eeee have thus the same form as the \eee for the motion of an elastic string with fixed end points under the influence of an external force. There are standard techniques for solving such equations. For example, one can find the desired result by expanding the solution in time $t$ in terms of eigenfunctions $\sin n\tr$  and expanding the ``external force" similarly (see \cite{CH}).
 
 To determine the ``dragging potential" $\om(t,\rh)$ we employ simpler \eeee $\RR_{02}=2\SS_{02}$ and $\RR_{12}=2\SS_{12}$, with $\RR_{02}$ and $\RR_{12}$ given by (\ref{66}) and (\ref{67}). The explicit solution of these \eeee with the source field $\ps$ representing specific ingoing and outgoing pulses of the rotating \cy \ggg waves demonstrates nicely the dragging effects of such waves. We treat this problem in-depth in paper \cite{LBK07}.

\sect{Dragging of inertial frames in axisymmetric \ssss with \cy waves and spinning string}

The vacuum metrics (\ref{47}) considered by Mashhoon {\it et al} \cite{MCQ}
as ``rotating gravitational waves" cannot have a regular axis since this requires the Killing orbits to admit orthogonal surfaces. However, the axis can represent a rotating cosmic string. Let us assume the `rotation parameter' $\om$ in   (\ref{47})  is small so that the terms in $\OO(\om^2)$ can be neglected. The inspection of vacuum field \eeee following from the Anzatz (\ref{47})  then reveals that one  can choose $W=\rh$ and \eeee for $\ps$ and $\ga$
 are the same as (\ref{29}) and (\ref{210}), see Appendix A. The rotational \ppp $\om$ is determined by the \eee
 \be
 R_{12}= - \fr{1}{2\rh}\lll  \rh^3 e^{-2\ga}\om' \rrr^{\bf .}=0,
 \lb{51}
 \ee
 and the constraint equation
 \be
 R_{02}=- \fr{1}{2\rh}\lll  \rh^3 e^{-2\ga}\om' \rrr'=0.
 \lb{52}
 \ee
 These two \eeee are directly related to the properties of the \am of \cy systems derived from the Komar expressions (\ref{41}) or  (\ref{43}) and  (\ref{46}). In  (\ref{46})
we indicated that one can evaluate the integral at $\rh\ra\inf$ since its value in vacuum does not depend on the location of the cylinder $\CC$. For the metrics of the form  (\ref{47})
 with small $\om$ we obtained the result  (\ref{49}) which again one can evaluate at $\rh\ra\inf$. However, it is evident from the constraint equation  (\ref{52}) that the same value for $J$
 results at any $\rh\ne0$ since $ \rh^3 e^{-2\ga}\om'$ is independent of $\rh$ and by (\ref{51}) independent of time. 
 
 Integration of (\ref{51}) and (\ref{52}) yields
 \be
 \om'=\fr{  e^{2\ga(t,\rh)}}{  \rh^3}\fr{\ka}{\pi}J,
 \lb{53}
 \ee
where the integration constant was chosen in accordance with the expression (\ref{49}) for the angular momentum. The \am is a small constant parameter with  $\fr{\ka J}{\pi}$ having the dimensions of   length. Hence,
\be
\om(t,\rh)= - \fr{\ka}{\pi}J\int^\inf_\rh\fr{e^{2\ga(t,\rh)}}{\rh^3}d\rh+\om_0(t);
\lb{54}
\ee
we put $\om_0=0$ because it can be transformed away by going to rotating axes, without a change of $J$ (cf. Section 4).

When no wave is present, (\ref{54}) gives 
$\om\eq\om_S= - \fr{\ka}{2\pi}\fr{J}{\rh^2}$;
the metric (\ref{47}) with $W=\rh$ becomes
precisely the metric (\ref{411}) of the massless spinning string
with \am $J$. The ``string perturbation" $\om_S$ causes rotation of local inertial frames at some finite $\rh$
as compared with frames at infinity. However, this relative rotation is time-independent as, e.g., in the Kerr metric.

Now consider $\ps$ and $\ga$ to represent an exact \cy \ggg wave, e.g., the Bonnor/Weber-Wheeler  wave given in (\ref{212}) and (\ref{213}). Since the rotational effects are given by a linear \ppp of the metric we can write the total \ppp (\ref{53}) as
$\om'=\om'_S +\om'_W$,
the time-independent $\om'_S$ is attributed to the spinning string, whereas the time-dependent part
\be
\om'_W=\fr{\ka J}{\pi\rh^3}(e^{2\ga(t,\rh)}-1)
\lb{58}
\ee
is associated with the presence of a \cy \ggg wave interacting with the string. 

Now at $t^2\gg a^2+\rh^2$, (\ref{213}) implies $\ga\thickapprox0$, so when the pulse is far away from the axis, the dragging near the axis is dominated by the string. As the wave proceeds towards the axis, the dragging due to its presence increases. The expansion of $\gamma$ at small $\tr$ [see (\ref{211})] reads 
\be
\ga=8b^2\tr^2\fr{\tt^2}{(1+\tt^2)^4}+\OO(\tr^4).
\lb{59}
\ee 
Hence, at small $\tr$ \eee (\ref{58}) gives
\be
\om'_W\sim\fr{16b^2\ka J}{\pi a^3} \fr{1}{\tr}\fr{ \tt^2}{(1+\tt^2)^4},
\lb{510}
\ee
so that
\be
\om_W\sim\fr{16b^2\ka  J}{\pi a^2} \ln\tr\fr{\tt^2}{(1+\tt^2)^4}.
\lb{511}
\ee
Since we assumed $\om$ - or, correspondingly, $J$ - small, our approximation will break down\footnote{A somewhat analogous situation arises in the problem of slowly rotating collapsing spherical shells when the shells approach the horizon \cite{KLB}.} at $\rh=0$, but it should be reliable at all $\rh$'s for which $\ka J/\pi\rh\ll 1$; this can be assured at any $\rh\ne0$ by choosing $J$ sufficiently small.

At spatial infinity, $t=$const., $\rh\ra\inf$, the function $\ga$ approaches a constant
\be
\ga=\ha b^2\LLL 1-\fr{2}{\tr^2}+\OO(\tr^{-4})  \RRR;
\lb{512}
\ee
the time-dependent terms appear only in $\OO(\tr^{-4})$. Substituting this expansion into  (\ref{58}) and integrating, we find the angular velocity of the rotation of inertial frames at large $\tr$:
\be
\om_W=\fr{\ka J}{\pi\rh^2}\lll  e^{b^2}-1  \rrr + \OO(\tr^{-4}).
\lb{513}
\ee
The \ppp $\om_S$ corresponding to the string also decays at infinity, so local inertial frames do not rotate with respect to the `fixed stars' at infinity, i.e., with respect to the `lines' $\vf=$const. However, they do rotate with respect to these lines close to the axis, at small $\tr$, with the time-dependent \av $\om_W$ given by (\ref{511}). This vanishes at times $|\tt|\gg1$ when the incoming or outgoing pulse is far away from the axis. It becomes maximal (${\dot\om}_W=0$) at $|\tt|=1/\sq{3}$. Then the pulse is close to the axis.

\vs 

\Large{\bf Acknowledgements}
 \vskip .5 cm
 \normalsize 
 \setlength{\baselineskip}{20pt plus2pt}
  
 J.B. and J.K. are grateful to the Institute of Astronomy, Cambridge University, and the Royal Society for hospitality and support.  J.B. also acknowledges the hospitality of the Albert Einstein Institute in Golm    and the Institute of Theoretical Physics at FSU in Jena, the support of the Alexander von Humbolt Foundation, the partial support from the Grant GA\v CR 202/06/0041 of the Czech Republic, of Grant No  LC06014 and MSM0021620860 of the Ministry of Education  and from SFB/TR7 in Jena.

\vskip .5 cm
\begin{appendix}
\setcounter{equation}{0}
\renewcommand{\theequation}{\Alph{section}.\arabic{equation}}
\section{ The Ricci tensor for the averaged rotating \ggg waves in 2+1 dimensional \sss}
\normalsize
\setlength{\baselineskip}{20pt plus2pt}
  \vs
Assume the metric is of the form (\ref{64}). The Ricci tensor components read as follows:
\ba
\RR_{00}&=&-\ddot\ga+\ga''-\fr{\ddot W}{W}+\fr{W'}{W}\ga'+\fr{\dot W}{W}\dot\ga\nn\\
&+&e^{-2\ga}W\lll 2\ga'W\om\om'+\ddot W\om^2 - W''\om^2 - 3W'\om\om' -\ts{\fr{1}{2}} W{\om'}^2 -W\om\om''-\ts{\fr{1}{2}} W^3\om^2{\om'}^2e^{-2\ga}     
       \rrr,\nn \\\nn
\\ \RR_{01}&= &\fr{W'}{W}\dot \ga - \fr{{\dot W}'}{W} + \fr{\dot W}{W}\ga'+e^{-2\ga} W\lll   W\dga \om\om' - \ts{\fr{1}{2}}W\om\dom' - {\ts\fr{3}{2}}\dot W\om \om'
\rrr,\nn\\
\RR_{02}&=&e^{-2\ga}W\lll  {\ts\fr{3}{2}}W'\om'+\ts{\fr{1}{2}}W\om''+W''\om - W\ga'\om'-\ddot W\om+\ts{\fr{1}{2}}W^3\om{\om'}^2 e^{-2\ga} \rrr,\nn\\
\RR_{11}&=&\ddot\ga-\ga''+\fr{W'}{W}\ga' - \fr{W''}{W}+\fr{\dW}{W}\dga+\ts{\fr{1}{2}} W^2{\om'}^2e^{-2\ga},\nn\\
\RR_{12}&=&\ts{\fr{1}{2}}e^{-2\ga}W\lll  W{\dom}'  +3\dW\om'-2W\dga\om' \rrr,\nn\\
\RR_{22}&=&e^{-2\ga}W\lll \ddot W - W'' -\ts{\fr{1}{2}}W^3{\om'}^2  e^{-2\ga}\rrr.\nn
 \ea
\end{appendix}

  \end{document}